\documentclass[12pt]{article}
\title{
\vspace{-8mm}
\rightline{\small{\tt hep-th/9902177}}
\vspace{-2mm}
\rightline{\small HUB--EP--99/14}
\vspace{-3mm}
\bf Monopole-like Excitations\\ 
as a Source of Confinement in the\\ 
SU(2)-Gluodynamics}
\author{Dmitri Antonov 
\thanks{E-mail address: 
{\tt antonov@mailbox.difi.unipi.it}}
\thanks{Permanent address: 
Institute of Theoretical and Experimenal Physics,
B. Cheremushkinskaya 25, RU-117 218 Moscow, Russia.}\\
{\it INFN-Sezione di Pisa, Universit\'a degli studi di Pisa,}\\ 
{\it Dipartimento di Fisica, 
Via Buonarroti, 2 - Ed. B - 56127 Pisa, Italy}\\
and\\ 
Dietmar Ebert \thanks{E-mail address: 
{\tt debert@physik.hu-berlin.de}}\\
{\it Institut f\"ur Physik, Humboldt-Universit\"at zu Berlin,}\\ 
{\it Invalidenstra{\ss}e 110, D-10115 Berlin, Germany}}
\date{}
\begin{document}
\maketitle

\vspace{1mm}
\centerline{\bf {Abstract}}
\vspace{3mm}
\noindent
By making use of the Abelian projection method, 
a dual version of the $SU(2)$-gluodynamics 
with manifest monopole-like excitations, arising from the 
integration over singular gauge transformations, is formulated 
in the continuum limit.
The resulting effective theory emerges due to the summation 
over the grand canonical ensemble of these excitations
in the dilute gas approximation.
As a result, the dual Abelian gauge boson acquires 
a nonvanishing (magnetic) mass due to the Debye screening effects 
in such a gas. 
The obtained theory 
is then used for the construction of 
the corresponding effective potential of monopole loop
currents and the string representation. 
Finally, by virtue of this 
representation, confining properties of the $SU(2)$-gluodynamics 
are emphasized.  

\vspace{5mm}

\newpage

\section{Introduction}

Nowadays, it is commonly argued that 
the method of Abelian projection~\cite{th} is one of the 
most challengable approaches to solve the problem of 
confinement in QCD 
within the dynamical scheme of a dual superconductor~\cite{2}
(for a recent review see {\it e.g.}~\cite{3}). 
In particular, a detailed perturbative analysis of the $SU(2)$-QCD within 
the dual approach has been performed~\cite{rein, kond}, and 
the asymptotic freedom of the resulting effective Abelian theory 
has been proved. As far as the property of confinement in the  
Abelian-projected $SU(N)$-QCD is concerned,
it has been argued in Refs.~\cite{kond, suz} 
that it occurs 
owing to the condensation of 
Cooper pairs of magnetic monopoles, 
described by the magnetic Higgs field. Spontaneous
breaking of the resulting $U(1)$ symmetries then 
leads to the generation of the 
mass terms of the dual gauge fields. This makes the effective
Abelian-projected $SU(N)$ gauge theory, obtained in this way, 
quite similar 
to the (London limit of the) dual Abelian Higgs type model
with the $[U(1)]^{N-1}$ gauge invariance. In the 
latter model, confinement can be analytically studied by casting the 
corresponding partition function into the form of an integral over 
the world-sheets of the closed Abrikosov-Nielsen-Olesen 
strings~\cite{ano} by making use of the so-called path-integral 
duality transformation. This transformation elaborated on 
for the Abelian Higgs model in Refs.~\cite{lee, polikarp}
has been employed for a derivation of the string representations 
for the partition functions and field strength correlators in 
Abelian-projected $SU(2)$- and $SU(3)$-QCD in Refs.~\cite{eur} 
and~\cite{more}, 
respectively~\footnote{The evaluation of field strength correlators 
in Abelian-projected theories by another methods has been 
independently performed in Ref.~\cite{bramb}.}.
After that, performing 
the derivative 
expansion~\cite{mpla} of the so-obtained string effective action, one  
gets as the first two terms of this expansion 
the usual Nambu-Goto term and the so-called 
rigidity term~\cite{rid}, whose coupling constants ensure 
confinement (in the sense of the Wilson's area law~\cite{wil})  
and stability of the Abrikosov-Nielsen-Olesen strings. 

The aim of the present paper 
is to derive an effective low-energy dual theory of Abelian-projected 
$SU(2)$-gluodynamics in the continuum limit 
by summing over the grand canonical ensemble 
of monopole loop currents, which emerge during the Abelian projection. 
Moreover, in this way we shall not make the standard 
assumption on the formation and subsequent condensation of Cooper 
pairs of magnetic monopoles, but shall rather treat the ensemble 
of monopole loop currents in the dilute gas approximation. Next, 
in order to achieve our main goal, which is a manifestation of 
confinement in the $SU(2)$-gluodynamics, we find it necessary 
to derive a string representation of the obtained theory. 
The latter one is implied as a certain mechanism realizing 
the independence of the Wilson loop describing a test particle, 
electrically charged {\it w.r.t.} the maximal Abelian $U(1)$ subgroup 
of the original $SU(2)$ group, of the shape of some surface 
bounded by the contour of this Wilson loop. The construction of 
such a mechanism, which will be performed below, is based on the 
summation over branches of the multivalued effective potential 
of monopole loop currents, which emerges in the representation 
of the obtained dual model in terms of an integral 
over these currents. Note that such an approach is the 4D generalization 
of the corresponding 3D one, investigated in Ref.~\cite{dec}.
In that paper, it has been demonstrated that this approach 
parallels the one proposed in Ref.~\cite{confstr} for the construction 
of a string representation of 3D compact QED (see also 
Ref.~\cite{dia} for the 4D generalizations). 
Notice also that within 
our approach, the 
dual gauge field acquires a mass dynamically, {\it i.e.}  
by virtue of the Debye screening 
in the gas of monopole loop currents. 
The appearance of this mass then leads to a 
nonvanishing string tension and thus confinement of an electrically 
charged test particle. 
Such a mechanism of the mass generation conceptually
differs from the one of Refs.~\cite{kond, fuc},  
which employed among others 
the cumulant 
expansion theorem in the bilocal approximation~\cite{bilocal}.
As a by-product of the 
present work, we propose a method of a derivation of the effective 
dual theory, describing a 4D dilute gas of monopole loop currents, 
which does not exploit  
the corresponding lattice partition function, as it has been 
done in Ref.~\cite{dia}.
 
The organization of the paper is as follows. In the next Section, 
we shall revisit a derivation of the effective dual theory, 
corresponding to the Abelian-projected $SU(2)$-gluodynamics.
After that, we shall perform the path-integral summation 
over the grand canonical ensemble of fluctuating random 
monopole loop currents, which emerge during the Abelian projection, 
in the dilute gas approximation and arrive at a certain effective 
field theory describing this ensemble.
In Section 3, this theory will be used for the calculation of the 
potential of monopole loop currents and the derivation of the 
corresponding string representation. Finally, the latter one will yield 
us confinement of an electrically charged test particle 
in the sense of the Wilson's area law. 
The r\'esum\'e 
of the work and concluding remarks are presented in Summary and 
Discussions.

\section{Effective Dual Theory of the 
Abelian-Projected $SU(2)$-Gluodynamics} 

In the present Section, we shall derive an effective dual model 
corresponding to the Abelian-projected $SU(2)$-gluodynamics. 
The starting points of this derivation will somewhat parallel 
that of Refs.~\cite{rein, kond, suz}. The action under 
study reads~\footnote{Throughout the present paper, we work in the 
Euclidean space-time.}

\begin{equation}
\label{1}
S_{\rm YM}\left[A_\mu^i\right]=\frac12{\rm tr}\int d^4x F_{\mu\nu}^2,
\end{equation}
where $F_{\mu\nu}=F_{\mu\nu}^iT^i$
with $F_{\mu\nu}^i=\partial_\mu A_\nu^i-\partial_\nu A_\mu^i+
g\varepsilon^{ijk}A_\mu^j A_\nu^k$ 
and $T^i=\frac{\tau^i}{2}$, $i=1,2,3$. Here, $\tau^i$'s stand for 
Pauli matrices, and $g$ is the QCD (``electric'') coupling constant. 

One can perform the gauge transformation
$A_\mu'=UA_\mu U^\dag+\frac{i}{g}U\partial_\mu U^\dag$, 
so that the gauge-transformed field $A_\mu'$ obeys the so-called 
maximal Abelian gauge fixing condition 
(see {\it e.g.} Refs.~\cite{rein, kond})
$\left(\partial_\mu\pm iga'_\mu\right)\left(A^{'1}_\mu\pm iA^{'2}_\mu
\right)=0$, 
where $a'_\mu\equiv A^{'3}_\mu$. 
Notice that the maximal Abelian gauge fixing condition 
can be written as follows 
${\cal D}_\mu^{'ab}A^{'b}_\mu=0$, where 
${\cal D}_\mu^{'ab}=\partial_\mu\delta^{ab}-g\varepsilon^{ab3}
a'_\mu$. Once being rewritten in this form, 
this gauge can be easily recognized as the standard 
background gauge~\cite{backgr}  
with the field $a'_\mu$ playing the r\^ole of the 
background~\footnote{Recently, 
in Ref.~\cite{ellw} this analogue between the two gauges 
has been employed for the investigation of the Wilsonian 
exact renormalization group flow of gluodynamics in the 
maximal Abelian gauge.}.
The gauge transformed field strength tensor then reads
$F'_{\mu\nu}=U\left(F_{\mu\nu}+F_{\mu\nu}^{\rm sing.}\right)U^\dag$,
where the singular contribution has the form 
$F_{\mu\nu}^{\rm sing.}=\frac{i}{g}\left(\left[\partial_\mu, 
\partial_\nu\right]U^\dag\right)U$.
This contribution comes about from the singular character 
of the matrix $U$ of the gauge transformation~\cite{th, 3, 
rein, kond, rein1} 
and describes world-sheets of the Dirac strings. 
Clearly, integration over all possible singular gauge transformations 
results to an integration over $F_{\mu\nu}^{\rm sing.}$.

Let us next single out the diagonal (neutral) component 
$a_\mu\equiv A_\mu^3$ of the field 
$A_\mu$ by making use of the decomposition~\footnote{From now on, 
we omit for brevity the prime denoting the gauge transformed 
fields, implying everywhere the maximal Abelian gauge fixing condition.} 
$A_\mu=a_\mu T^3+A_\mu^aT^a\equiv {\cal A}_\mu+{\cal C}_\mu$,  
where $a=1,2$. Consequently, 
one has for the field strength tensor 

\begin{equation}
\label{2}
F_{\mu\nu}\equiv F_{\mu\nu}\left[{\cal A}+{\cal C}\right]=
F_{\mu\nu}\left[{\cal A}\right]+\left(D\left[{\cal A}\right]\wedge{\cal C}
\right)_{\mu\nu}-ig\left[{\cal C}_\mu,{\cal C}_\nu\right], 
\end{equation}
where $\left({\cal O}\wedge{\cal G}\right)_{\mu\nu}\equiv {\cal O}_\mu 
{\cal G}_\nu-{\cal O}_\nu{\cal G}_\mu$, and  
$D_\mu\left[{\cal A}\right]
=\partial_\mu-ig\left[{\cal A}_\mu,\cdot\right]$.  
Eq.~(\ref{2}) can be straightforwardly rewritten as follows 
$F_{\mu\nu}=\left(f_{\mu\nu}+C_{\mu\nu}\right)T^3+S_{\mu\nu}^aT^a$. 
Here, 
$f_{\mu\nu}=(\partial\wedge a)_{\mu\nu}$ and  
$C_{\mu\nu}=g\varepsilon^{ab3}A_\mu^a A_\nu^b$ stand for the 
contributions of diagonal and off-diagonal components of the gluon field 
to the diagonal part of the field strength tensor, respectively, and 
$S_{\mu\nu}^a=\left({\cal D}^{ab}\wedge A^b\right)_{\mu\nu}$ is the 
off-diagonal part of the field strength tensor.
This yields the following decomposition of the 
action~(\ref{1}) 
(taken now on the gauge transformed fields)  

\begin{equation}
\label{3}
S_{\rm YM}\left[A^i_\mu\right]=
\frac14\int d^4x\left(f_{\mu\nu}+C_{\mu\nu}+\left(F_{\mu\nu}^{\rm 
sing.}\right)^3\right)^2 
+\frac14\int d^4x \left(S_{\mu\nu}^a+\left(F_{\mu\nu}^{\rm sing.}
\right)^a\right)^2,
\end{equation}
where $\left(F_{\mu\nu}^{\rm sing.}
\right)^i=2{\,}{\rm tr}{\,}\left(T^iF_{\mu\nu}^{\rm sing.}\right)$.
It is worth remarking that the non-Abelian commutator term 
$C_{\mu\nu}$, when evaluated 
with the singular part of the 
gauge transformed field, 
$A_\mu^{\rm sing.}=\frac{i}{g}U\partial_\mu U^\dag$,  
generates among others monopole contributions.
Such monopole terms have, however, been shown to become cancelled by the  
corresponding terms 
arising during the evaluation of the Abelian field strength 
$f_{\mu\nu}$ at 
$\left(A_\mu^{\rm sing.}\right)^3$~\cite{kond, rein1}. This leaves 
in Eq.~(\ref{3}), besides the contributions of the 
non-singular gauge field 
configurations to be treated as quantum fluctuations, only the (singular) 
contributions of Dirac strings.

As it has been demonstrated in Refs.~\cite{rein, kond}, 
all the terms on the R.H.S. of Eq.~(\ref{3}) 
depending on the off-diagonal gluons $A_\mu^a$'s 
contribute 
to the momentum dependence of the running coupling constant and 
yield asymptotic freedom. 
Namely, the running 
coupling constant coincides with that of the original gluodynamics and 
reads 
$g(\mu)^{-2}=g(\mu_0)^{-2}+
\frac{b_0}{8\pi^2}\ln\frac{\mu}{\mu_0}$, where  
$b_0=\frac{11 C_2(G)}{3}$
with $C_2(G)$ standing for the Casimir operator of the adjoint 
representation of the group $G=SU(2)$ 
under consideration, {\it i.e.} 
$C_2(G)=2$. Since in what follows our aim 
will be the 
investigation of the confining ({\it i.e.} infrared) 
properties of the Abelian-projected 
$SU(2)$-gluodynamics (rather than the problems of its renormalization, 
related 
to the region of asymptotic freedom), we 
shall disregard the $A_\mu^a$-dependent terms (This approximation 
is usually referred to as the Abelian dominance 
hypothesis~\cite{abdom}.). Within this approximation, the resulting 
effective action takes the form

\begin{equation}
\label{5}
S_{\rm eff.}\left[a_\mu, {\cal F}_{\mu\nu}\right]=
\frac14\int d^4x\left(f_{\mu\nu}+
{\cal F}_{\mu\nu}
\right)^2, 
\end{equation}
where we have denoted for brevity ${\cal F}_{\mu\nu}\equiv 
\left(F_{\mu\nu}^{\rm sing.}\right)^3$. 
 
The monopole current is defined 
via the 
modified Bianchi identities as follows 

\begin{equation}
\label{magcur}
j_\nu^M=\partial_\mu\left(\tilde f_{\mu\nu}+
\tilde{\cal F}_{\mu\nu}
\right)=\frac12\varepsilon_{\mu\nu\lambda\rho}\partial_\mu 
{\cal F}_{\lambda\rho} 
\end{equation}
with $\tilde f_{\mu\nu}=\frac12\varepsilon_{\mu\nu\lambda\rho}
f_{\lambda\rho}$, {\it etc}. Thus, in what follows we shall regard the 
obtained effective theory~(\ref{5}) as a $U(1)$ gauge theory with  
monopole loop currents. 
Our aim then will 
be to investigate confining properties of such a theory by a derivation 
of its string representation.
To this end, let us first cast 
the partition function under study, 
${\cal Z}=\int {\cal D}{\cal F}_{\mu\nu}
{\cal D}a_\mu\exp\left(-S_{\rm eff.}\left[a_\mu, {\cal F}_{\mu\nu}
\right]\right)$,  
to the dual 
form~\footnote{Notice that the gauge fixing term of the Abelian 
field is  
assumed to be included into the integration measure 
${\cal D}a_\mu$.}. This can be done by making use of the 
first-order formalism, {\it i.e.}  
linearizing the square $f_{\mu\nu}^2$ in 
Eq.~(\ref{5}) by introducing an integration over an auxiliary 
antisymmetric tensor field $b_{\mu\nu}$ as follows 

\begin{equation}
\label{6}
{\cal Z}=\int {\cal D}{\cal F}_{\mu\nu}{\cal D}a_\mu 
{\cal D}b_{\mu\nu}\exp\left\{-\int d^4x\left[\frac14
b_{\mu\nu}^2+\frac{i}{2}\tilde b_{\mu\nu} 
f_{\mu\nu}+\frac12f_{\mu\nu}{\cal F}_{\mu\nu}+
\frac14{\cal F}_{\mu\nu}^2
\right]\right\}.
\end{equation}
Integration over the $a_\mu$-field leads to the constraint $\partial_\mu
\left(\tilde b_{\mu\nu}-i{\cal F}_{\mu\nu}\right)=0$, 
whose resolution yields $b_{\mu\nu}=i\tilde{\cal F}_{\mu\nu}+
(\partial\wedge b)_{\mu\nu}$, where $b_\mu$ is now the 
``magnetic'' potential dual to the ``electric'' potential $a_\mu$. 
Substituting 
this representation for $b_{\mu\nu}$ into Eq.~(\ref{6}), we get

\begin{equation}
\label{7}
{\cal Z}=
\left<\int 
{\cal D}b_\mu\exp\left\{-\int d^4x\left[\frac14 b_{\mu\nu}^2-i 
b_\mu j_\mu^M\right]\right\}\right>_{j_\mu^M},
\end{equation}
where from now on, $b_{\mu\nu}$ denotes simply $(\partial\wedge 
b)_{\mu\nu}$. In Eq.~(\ref{7}), the integration over
${\cal F}_{\mu\nu}$'s has transformed to a certain average over monopole  
loop currents, 
$\left<\ldots\right>_{j_\mu^M}$, whose concrete form will be 
specified below. 

It is worth noting that due to the conservation of the monopole 
current $j_\mu^M$, the dual action standing in the exponent 
on the R.H.S. of Eq.~(\ref{7}) is invariant under the 
magnetic gauge 
transformations $b_\mu\to b_\mu+\partial_\mu\chi$. Again, we shall 
imply that the gauge fixing term for the $b_\mu$-field is included into 
the integration measure ${\cal D}b_\mu$. Moreover, we shall specify the 
gauge to be the Fock-Schwinger one, {\it i.e.} $x_\mu b_\mu(x)=0$. 

Our next aim is to sum up over the 
ensemble of monopole loop currents in the dual 
theory~(\ref{7}). 
To this end, we shall treat this ensemble as the grand canonical one and 
make an assumption 
that monopole loop currents form a dilute gas. Then,
since the energy of a single monopole is known to be a quadratic 
function of its flux, it is more 
energetically favorable for the vacuum to 
support a configuration of two monopoles of a unit magnetic charge 
than one monopole of the double charge. Therefore, 
only the monopoles with the minimal 
charges $q_a g_m$ with $q_a=\pm 1$ are essential, whereas the 
ones with $\left|q_a\right|>1$ tend to dissociate into  
those with $\left|q_a\right|=1$. Here, the magnetic 
coupling constant $g_m$ is related to the QCD coupling $g$ via the 
topological 
quantization condition $gg_m=4\pi n$. In what follows, we shall 
set in this condition $n=1$, 
which parallels the above restriction to the monopoles 
possessing the 
minimal charge only. Obviously, the same restriction then holds for 
the Dirac strings ending up at monopole-antimonopole pairs, as well. 
The collective current 
of $N$ monopoles takes the form

\begin{equation}
\label{8}
j_\mu^{M{\,}(N)}(x)=\frac{4\pi}{g}\sum\limits_{a=1}^{N}q_a\oint dz_\mu^a
\delta\left(x-x^a(\tau)\right),
\end{equation}
where the $a$-th monopole loop current 
is parametrized by the vector $x_\mu^a(\tau)=y_\mu^a+
z_\mu^a(\tau)$, $0\le\tau\le 1$. Here, 
$y_\mu^a=\int\limits_{0}^{1}d\tau x_\mu^a(\tau)$ 
denotes the position of the $a$-th loop current, whereas the vector 
$z_\mu^a(\tau)$ corresponds to its shape, both of which should be averaged 
over independently in 
$\left<\ldots\right>_{j_\mu^M}$~\footnote{As it follows  
from Eq.~(\ref{magcur}), ${\cal F}_{\mu\nu}$ corresponding to the 
current~(\ref{8}) is 
nothing else, but the field strength tensor of $N$ Dirac 
strings, ${\cal F}_{\mu\nu}(x)=-\frac{4\pi}{g}\frac12
\varepsilon_{\mu\nu\lambda\rho}
\sum\limits_{a=1}^{N}q_a\int d\sigma_{\lambda\rho}\left(x^a(\xi)\right)
\delta\left(x-x^a(\xi)\right)$. Here, $x^a(\xi)$ is a vector parametrizing 
the world-sheet of the $a$-th string with $\xi$ standing for the 
two-dimensional coordinate.}. Namely, the average $\left<\ldots
\right>_{j_\mu^M}$ with the collective current~(\ref{8}) takes the form

\begin{equation}
\label{aver}
\left<{\cal O}\left[j_\mu^M\right]\right>_{j_\mu^M}=
\prod\limits_{i=1}^{N}\int d^4y^i {\cal D}z^i\mu\left[z^i\right]
\sum\limits_{q_a=\pm 1}^{}{\cal O}\left[j_\mu^{M{\,}(N)}\right].
\end{equation}
Here, $\mu\left[z^i\right]$ is a certain rotation- and 
translation invariant integration measure over the shapes of monopole 
loop currents,  
whose concrete form will not be specified 
here (For example, one can take it in the form of the properly normalized 
measure of an ensemble of oriented random loops, representing 
trajectories of scalar particles, 

$$\int {\cal D}z^i\mu\left[z^i\right]{\cal O}\left[z^i\right]=
{\cal N} \int\limits_{0}^{+\infty}\frac{ds_i}{s_i}
\int\limits_{u\left(0\right)=u\left(s_i\right)}^{}
{\cal D}u\left(s_i'\right)
\exp\left(-\frac14\int\limits_{0}^{s_i}\dot u^2\left(s_i'\right)
ds_i'\right){\cal O}\left[u\left(s_i'\right)\right],$$
where the vector $u_\mu\left(s_i'\right)$ parametrizes the same 
contour as the vector $z_\mu^i(\tau)$.). 

One can now write down the contribution of $N$ monopole loop currents 
to the partition function of their grand canonical ensemble. 
Owing to Eqs.~(\ref{8}) and~(\ref{aver}) it reads 

$$
{\cal Z}^M\left[b_\mu\right]=
1+\sum\limits_{N=1}^{\infty}\frac{\zeta^N}{N!}\left<
\exp\left(i\int d^4x b_\mu j_\mu^M\right)\right>_{j_\mu^M}=$$

\begin{equation}
\label{9}
=1+\sum\limits_{N=1}^{\infty}\frac{\left(2\zeta\right)^N}{N!}
\left\{\int d^4y\int {\cal D}z\mu[z]\cos\left(\frac{4\pi}{g}
\oint dz_\mu b_\mu(x)\right)\right\}^N.
\end{equation}
Here, 
$\zeta\propto {\rm e}^{-S_0}$ is the so-called fugacity term (Boltzmann 
factor of a single monopole loop current) 
of dimension $({\rm mass})^4$ with the action of a single 
loop current given by 
$S_0={\rm const.} g_m^2$.

In order to evaluate the path-integral over $z_\mu$'s in Eq.~(\ref{9}), 
let us employ
the above mentioned dilute gas approximation, which 
requires 
that typical distances between monopole loop currents are much larger than 
their sizes. This means that generally $\left|y^a\right|\gg\left|z^a
\right|$, where from now on $\left|y\right|\equiv\sqrt{y_\mu^2}$. 
Let us denote characteristic distances $\left|y\right|$  
by $L$, characteristic sizes of monopole loop currents 
$\left(=\int\limits_{0}^{1}d\tau 
\sqrt{\dot z^2}\right)$ by $a$, and perform the Taylor expansion 
of $b_\mu(x)$ up to the first order in $a/L$ (which is the first one 
yielding a nonvanishing contribution to the integral 
$\oint dz_\mu b_\mu(x)$   
on the R.H.S. of Eq.~(\ref{9})), 

\begin{equation}
\label{10}
b_\mu(x)=b_\mu(y)+L^{-1}z_\nu n_\nu b_\mu(y)+O\left(\left(\frac{a}{L}
\right)^2\right).
\end{equation}
Here, we have denoted $n_\nu=\frac{y_\nu}{\left|y\right|}$ and estimated 
the derivative $\partial/\partial y_\nu$ as $n_\nu/L$. Then, 
the substitution of 
expansion~(\ref{10}) into Eq.~(\ref{9}) yields 

$$
\int {\cal D}z\mu[z]\cos\left(\frac{4\pi}{g}
\oint dz_\mu b_\mu(x)\right)\simeq\int {\cal D}z\mu[z] 
\cos\left(\frac{4\pi}{gL}n_\nu b_\mu(y){\cal P}_{\mu\nu}[z]\right)=$$

\begin{equation}
\label{11}
=\sum\limits_{n=0}^{\infty}\frac{(-1)^n}{(2n)!}\left(\frac{4\pi}{gL}
\right)^{2n}n_{\nu_1}b_{\mu_1}(y)\cdots n_{\nu_{2n}}b_{\mu_{2n}}(y)
\int {\cal D}z\mu[z]{\cal P}_{\mu_1\nu_1}[z]\cdots 
{\cal P}_{\mu_{2n}\nu_{2n}}[z],
\end{equation}
where 
${\cal P}_{\mu\nu}[z]\equiv\oint dz_\mu z_\nu$ 
stands for the tensor area associated with the contour parametrized by 
$z_\mu(\tau)$~\footnote{One can check that for the plane contour, 
${\cal P}_{\mu\nu}=-{\cal P}_{\nu\mu}=-S$, 
$\mu<\nu$, where $S$ is the area inside 
the contour.}. Due to the rotation- and 
translation invariance of the measure $\mu[z]$, the average of the 
product of the tensor areas can be written in the form 

\begin{equation}
\label{12}
\int {\cal D}z\mu[z]{\cal P}_{\mu_1\nu_1}[z]\cdots 
{\cal P}_{\mu_{2n}\nu_{2n}}[z]=
\frac{\left(a^2\right)^{2n}}{(2n-1)!!}\left[\hat 1_{\mu_1\nu_1, 
\mu_2\nu_2}\cdots \hat 1_{\mu_{2n-1}\nu_{2n-1},\mu_{2n}\nu_{2n}}+{\,} 
{\rm permutations}{\,}\right].
\end{equation}
Here  
$\hat 1_{\mu\nu,\lambda\rho}=\frac12\left(\delta_{\mu\lambda}
\delta_{\nu\rho}-\delta_{\mu\rho}\delta_{\nu\lambda}\right)$, 
and the normalization factor $(2n-1)!!$ is explicitly extracted 
out since the sum in square brackets on the R.H.S. of Eq.~(\ref{12}) 
contains $(2n-1)!!$ terms. Substituting now Eq.~(\ref{12}) into 
Eq.~(\ref{11}), recalling that we have adopted for 
the field $b_\mu$ the Fock-Schwinger gauge, so that 
$n_\mu b_\mu(y)=0$~\footnote{Within the dilute gas approximation, 
where $\frac{\partial}{\partial y_\nu}\to\frac{n_\nu}{L}$, 
the Fock-Schwinger 
gauge is equivalent to the Lorentz one.},  
and denoting $\frac{2\sqrt{2}\pi a^2}{gL}\left(\ll a\right)$ by 
$\Lambda^{-1}$, where $\Lambda$ 
acts as a natural UV momentum cutoff,  
we finally obtain 

$$
\int {\cal D}z\mu[z]\cos\left(\frac{4\pi}{g}
\oint dz_\mu b_\mu(x)\right)\simeq
\cos\left(\frac{\left|b_\mu(y)\right|}{\Lambda}\right).$$
Owing to this result, the 
partition function of the grand canonical ensemble of 
monopole loop currents reads ${\cal Z}^M\left[b_\mu\right]=\exp\left[
2\zeta\int d^4x\cos\left(\frac{\left|b_\mu\right|}{\Lambda}\right)
\right]$. Together with Eq.~(\ref{7}), it yields the desired 
expression for the partition function of an effective dual theory 
of the Abelian-projected $SU(2)$-gluodynamics, which has the form

\begin{equation}
\label{13}
{\cal Z}=
\int {\cal D}b_\mu\exp\left\{-\int d^4x\left[
\frac14 b_{\mu\nu}^2-2\zeta\cos\left(
\frac{\left|b_\mu\right|}{\Lambda}\right)
\right]\right\}.
\end{equation}
The (``magnetic'') 
Debye mass of the $b_\mu$-field, which it acquires due to the screening 
by magnetic loop currents, can now be immediately read off
from the expansion of the cosine and has the form 
$m=\frac{\sqrt{2\zeta}}{\Lambda}$. Notice also that 
a partition function of the type~(\ref{13}) (considered {\it ad~hoc} as 
a continuum version of the corresponding lattice expression) 
has been used  
in Ref.~\cite{dia} as a starting point for the construction 
of the string representation of the 4D compact QED. 
Our construction of an analogous representation for the model~(\ref{13}) 
will be performed in a more simple way. Namely, we shall construct 
such a string representation by virtue of the representation of the 
model under study in terms of the monopole loop currents, which is a 
4D generalization of the corresponding expression for the 3D 
partition function in terms of the monopole densities, investigated 
in Ref.~\cite{dec}.

\section{String Representation and Confinement}

In the present Section, we shall construct 
the string representation for the Wilson loop of an electrically charged 
({\it w.r.t.} the maximal Abelian $U(1)$ subgroup of the original 
$SU(2)$ group) test 
particle in the effective Abelian-projected 
theory~(\ref{13}). Such a representation will enable us to 
manifest confinement in this theory. 
To get the desired 
string representation, we shall first derive the 
representation of the corresponding partition function in 
terms of the integral over monopole loop currents. 
To this end, notice that  
integrating over the field $b_\mu$ 
in Eq.~(\ref{7}) one gets for the statistical weight 
$Z\left[j_\mu^M\right]$ defined by the relation
${\cal Z}\equiv\left<Z\left[j_\mu^M\right]
\right>_{j_\mu^M}$ 
an expression in the form of the Coulomb 
interaction between the monopole loop currents, 

$$
Z\left[j_\mu^M\right]=\exp\left(
-\frac{1}{8\pi^2}\int d^4xd^4x' j_\mu^M(x)
\frac{1}{(x-x')^2}j_\mu^M(x')\right).$$
Owing to this equation, 
one has~\footnote{
In the thermodynamic limit, 
where the number of monopole loop currents $N$
and the 
four-volume of observation $V$
infinitely increase  
with the density of the loop currents $\rho=N/V$ being kept fixed,
the collective current~(\ref{8}) can be treated
as a continuous function of disorder type. In what follows, we assume
that the ``free path length'' $L$ of monopole loop currents  
is much smaller than the 
characteristic size of the  Wilson loop of an external electrically 
charged test particle.}

$$
{\cal Z}=1+\sum\limits_{N=1}^{\infty}\frac{\zeta^N}{N!}
\left<\int {\cal D}j_\mu\delta\left(j_\mu-j_\mu^M\right)
Z\left[j_\mu\right]\right>_{j_\mu^M}=
$$

\begin{equation}
\label{14}
=\int {\cal D}j_\mu {\cal D}\lambda_\mu\exp\left[
-\frac{1}{8\pi^2}
\int d^4x d^4x' j_\mu(x)\frac{1}{(x-x')^2}
j_\mu(x')-i\int d^4x \lambda_\mu j_\mu+
2\zeta\int d^4x\cos\left(\frac{\left|\lambda_\mu\right|}{\Lambda}
\right)\right],
\end{equation}
where the term fixing the Fock-Schwinger gauge for the 
Lagrange multiplier $\lambda_\mu$ is again assumed to be included 
into the integration measure. Notice that ${\cal D}j_\mu$ 
here is the standard 
integration measure over the vector field, 
which is of the same form as ${\cal D}\lambda_\mu$.  

Clearly, due to the $\delta$-function standing on the 
R.H.S. of the first equality in Eq.~(\ref{14}), if we integrate 
the Lagrange multiplier $\lambda_\mu$ out of this equation, the 
resulting expression will be just the desired representation 
of the partition function in terms of the monopole loop 
currents. In order to carry out such an integration,
one should solve the saddle-point equation

\begin{equation}
\label{15}
\frac{\lambda_\mu}{\left|\lambda_\mu\right|}
\sin\left(\frac{\left|\lambda_\mu\right|}{\Lambda}\right)=
-\frac{i\Lambda}{2\zeta} 
j_\mu.
\end{equation} 
This can be done by noting that its L.H.S. is a vector in the 
direction $\lambda_\mu$, which means that it can be equal to the 
R.H.S. only provided that 
the direction of the vector $\lambda_\mu$ coincides 
with the direction of the vector $j_\mu$. Therefore, it is reasonable 
to seek for a solution to Eq.~(\ref{15}) 
in the form $\lambda_\mu=\left|\lambda_\mu\right|
\frac{j_\mu}{\left|j_\mu\right|}$. Then, 
Eq.~(\ref{15}) reduces to the scalar 
equation $\sin\left(\frac{\left|\lambda_\mu\right|}{\Lambda}\right)=
-\frac{i\Lambda\left|j_\mu\right|}{2\zeta}$. Straightforward 
solution of the 
latter one yields the desired representation for the partition 
function

\begin{equation}
\label{16}
{\cal Z}=\int Dj_\mu\exp\left\{-\left[
\frac{1}{8\pi^2}    
\int d^4x d^4x' j_\mu(x)\frac{1}{(x-x')^2}
j_\mu(x')+V\left[j_\mu\right]\right]\right\},
\end{equation}
where the complex-valued effective potential of monopole loop 
currents reads 

$$
V\left[j_\mu\right]=$$

\begin{equation}
\label{pot}
=\sum\limits_{n=-\infty}^{+\infty}
\int d^4x\left\{\Lambda \left|j_\mu\right|\left[\ln\left[
\frac{\Lambda}{2\zeta}\left|
j_\mu\right|+\sqrt{1+\left(
\frac{\Lambda}{2\zeta}\left|
j_\mu\right|\right)^2}\right]+2\pi in\right]-2\zeta
\sqrt{1+\left(
\frac{\Lambda}{2\zeta}\left|
j_\mu\right|\right)^2}
\right\}.
\end{equation}

Let us now proceed with the string representation of the Wilson loop 
in the effective theory~(\ref{13}).
Assuming for a while that the monopole loop currents are absent,  
one has for this 
object the following expression 

$$\left<W(C)\right>_{a_\mu}=\left<\frac12{\rm tr}{\,}P\exp
\left(ig\oint\limits_{C}^{}dx_\mu a_\mu T^3\right)\right>_{a_\mu},~ 
{\rm where}~ 
\left<\ldots\right>_{a_\mu}=\frac{\int {\cal D}a_\mu\left(\ldots\right)
\exp\left(-\frac14\int d^4x f_{\mu\nu}^2\right)}{\int {\cal D}a_\mu
\exp\left(-\frac14\int d^4x f_{\mu\nu}^2\right)}.$$ 
Next, the $P$-ordering can be omitted, since all the matrices commute 
with each other, after which we obtain

$$\left<W(C)\right>_{a_\mu}=\left<\cos\left(\frac{g}{2}\oint\limits_{C}^{}
dx_\mu a_\mu\right)\right>_{a_\mu}=\left<\exp\left(\frac{ig}{2}
\oint\limits_{C}^{}dx_\mu a_\mu\right)\right>_{a_\mu}=$$

$$
=\exp\left(-\frac{g^2}{32\pi^2}\oint
\limits_{C}^{}dx_\mu\oint\limits_{C}^{}dy_\mu\frac{1}{(x-y)^2}\right),$$
which is the standard ``perimeter'' (Gaussian) 
contribution to the Wilson loop.

However in the presence of monopole loop currents, one should 
properly extend the field strength tensor $f_{\mu\nu}$
in analogue to 
Eqs.~(\ref{5}) and (\ref{magcur})
in order to satisfy Bianchi identities modified by the current 
$j_\mu$. 
This can be done by 
using the  
complete field strength tensor 
$f_{\mu\nu}+h_{\mu\nu}$, where    
the fluctuating antisymmetric tensor-disorder field $h_{\mu\nu}$ 
(the so-called Kalb-Ramond field~\cite{KR}) just obeys these modified 
identities, {\it i.e.} $\partial_\mu\tilde h_{\mu\nu}=j_\nu$. 

By virtue of 
the Stokes theorem, 
we then obtain for the full Wilson loop the following expression

\begin{equation}
\label{17}
\left<W(C)\right>=\left<\exp\left[\frac{ig}{4}\int\limits_{\Sigma}^{}
d\sigma_{\mu\nu}\left(f_{\mu\nu}+h_{\mu\nu}\right)\right]
\right>_{a_\mu, j_\mu}=
\left<W(C)\right>_{a_\mu}\left<\exp\left(\frac{ig}{4}
\int\limits_{\Sigma}^{} 
d\sigma_{\mu\nu}h_{\mu\nu}\right)\right>_{j_\mu}. 
\end{equation}
Here, the average over currents is defined by the partition 
function~(\ref{16}), and 
$\Sigma$ is an arbitrary surface bounded by 
the contour $C$.
Expressing 
$h_{\mu\nu}$ via $j_\mu$, we can rewrite the last average on the 
R.H.S. of Eq.~(\ref{17}) directly as 

\begin{equation}
\label{18}
\left<\exp\left(-\frac{ig}{2}\int d^4x j_\mu\eta_\mu
\right)\right>_{j_\mu}.
\end{equation} 
Here,

\begin{equation}
\label{solid}
\eta_\mu(x)=\frac{1}{8\pi^2}\varepsilon_{\mu\nu\lambda\rho}
\frac{\partial}{\partial x_\nu}\int\limits_{\Sigma}^{}
d\sigma_{\lambda\rho}(x(\xi))\frac{1}{(x-x(\xi))^2}
\end{equation}
stands for the 4D solid angle, under which the surface $\Sigma$  
shows up to an observer located at the 
point $x$ with $\xi=\left(\xi^1, \xi^2\right)$ denoting the 
2D coordinate 
(If $\Sigma$ is a closed surface surrounding the point $x$  
than by virtue of the Gauss law, 
$d\tilde\sigma_{\mu\nu}\to dS_\mu\partial_\nu-dS_\nu\partial_\mu$, 
one can check that 
for the conserved current $j_\mu$, $\int d^4x j_\mu\eta_\mu=
\int dS_\mu j_\mu$, as it should be. Here, $dS_\mu$ stands for the 
oriented element of the hypersurface bounded by $\Sigma$.). 
An apparent $\Sigma$-dependence 
of Eq.~(\ref{18}) actually drops out due to the summation over branches 
of the multivalued potential~(\ref{pot}). This is the essence 
of the string representation of the Wilson loop in the effective 
dual theory~(\ref{13}).

Let us now consider the weak-field limit, {\it i.e.}  
the limit $\Lambda\left|j_\mu\right|\ll\zeta$, and investigate the 
(stable) minimum of the real branch of the potential of monopole 
loop currents. 
This corresponds to 
extracting the term with $n=0$ from the whole sum in Eq.~(\ref{pot}). 
Then, since we have 
restricted ourselves to the only one branch of the potential, the  
$\Sigma$-independence of the Wilson loop is spoiled. In order to restore it, 
let us choose $\Sigma$ to be the surface of the minimal area for a given 
contour $C$, unambiguously defined by this contour (see discussion in 
Ref.~\cite{dec}), 
$\Sigma=\Sigma_{\rm min.}\left[C\right]$. Then the Wilson 
loop takes the form 

$$
\left<W(C)\right>_{\rm weak-field}=\left<W(C)\right>_{a_\mu}\times
$$

\begin{equation}
\label{weak}
\times\int {\cal D}j_\mu\exp\left\{-\left[\frac{1}{8\pi^2} 
\int d^4x d^4x' j_\mu(x)\frac{1}{(x-x')^2}
j_\mu(x')+\frac{\Lambda^2}{4\zeta}\int d^4xj_\mu^2+
\frac{ig}{2}\int d^4x j_\mu
\eta_\mu\right]\right\}, 
\end{equation}
where now $\eta_\mu$ is defined by Eq.~(\ref{solid}) with the replacement 
$\Sigma\to\Sigma_{\rm min.}$. 
Recalling the expression for $j_\mu$ via $h_{\mu\nu}$, Eq.~(\ref{weak}) 
can be written as follows

\begin{equation}
\label{19}
\left<W(C)\right>_{\rm weak-field}=\left<W(C)\right>_{a_\mu}
\int {\cal D}h_{\mu\nu}\exp\left[-\int d^4x\left(\frac{\Lambda^2}{24\zeta}
H_{\mu\nu\lambda}^2+\frac14 h_{\mu\nu}^2\right)+\frac{ig}{4}
\int\limits_{\Sigma_{\rm min.}}^{} 
d\sigma_{\mu\nu}h_{\mu\nu}\right],
\end{equation}
where $H_{\mu\nu\lambda}=\partial_\mu h_{\nu\lambda}+\partial_\lambda
h_{\mu\nu}+\partial_\nu h_{\lambda\mu}$ is the field strength tensor 
of the Kalb-Ramond field $h_{\mu\nu}$. 
It is worth noting, that the mass of the Kalb-Ramond field following 
from the quadratic part of the action standing in the exponent 
on the R.H.S. of 
Eq.~(\ref{19}) is equal to the Debye mass $m$ of the field $b_\mu$
following from Eq.~(\ref{13}). Integration over the Kalb-Ramond 
field  
is now straightforward and can be performed along the lines of 
Ref.~\cite{eur}. Obviously, after such an integration, one gets 
the string effective action 
$S_{\rm str.}=-\ln\left<W(C)\right>_{\rm weak-field}$ 
in the form of 
an interaction between two world-sheet elements mediated by the 
propagator of this field. A certain part of this interaction can be 
rewritten by the Stokes theorem as the ``perimeter'' Yukawa type 
interaction (see Ref.~\cite{eur} for details).  
The remaining part, once being 
expanded 
in powers of the derivatives {\it w.r.t.} $\xi^a$'s 
(which is equivalent to the $1/m$-expansion) by virtue of the 
results of Ref.~\cite{mpla}, 
yields as the first two terms of this expansion the standard 
Nambu-Goto one and the so-called rigidity term~\cite{rid}, {\it i.e.}  

\begin{equation}
\label{streffact}
S_{\rm str.}\simeq 
\sigma\int d^2\xi\sqrt{\hat g}+\frac{1}{\alpha_0}
\int d^2\xi\sqrt{\hat g}\hat g^{ab}\left(\partial_a t_{\mu\nu}\right)
\left(\partial_b t_{\mu\nu}\right).
\end{equation}
Here, 
$\partial_a=\partial/\partial\xi^a$, 
$\hat g=\det \left|\left|\hat g^{ab}\right|\right|$ 
with $\hat g^{ab}=(\partial^a x_\mu(\xi))
(\partial^b x_\mu(\xi))$ being the induced metric tensor of the 
world-sheet, and $t_{\mu\nu}=\frac{\varepsilon^{ab}}{\sqrt{\hat g}}
\left(\partial_a x_\mu(\xi)\right)\left(\partial_b x_\nu(\xi)\right)$ 
standing for the so-called extrinsic curvature tensor. The string tension 
$\sigma$ of the Nambu-Goto term ({\it i.e.} the coefficient in the Wilson's 
area law) 
and the inverse coupling constant of the rigidity term, 
$1/\alpha_0$, are completely determined via the parameters of 
the model~(\ref{13}) and read 
$\sigma\simeq
\frac{g^2\zeta}{8\pi\Lambda^2}\ln\frac{1}{c}$ and 
$\frac{1}{\alpha_0}=-\frac{g^2}{128 \pi}$. 
Here, $c$ stands for a characteristic small 
dimensionless parameter, which in the model under study is reasonable 
to be set 
$c\sim g\zeta^{1/4}/\Lambda$.
Notice that the string tension 
is obviously proportional to the square of the 
Debye mass $m$ of the dual gauge field 
$b_\mu$ and consequently nonanalytic in the QCD coupling constant $g$. 
This result reflects the nonperturbative nature of confinement 
in the effective Abelian-projected theory~(\ref{13})  
similar to that in the original non-Abelian $SU(2)$-gluodynamics.

In conclusion of this Section, note that the signs of the string tension 
and coupling constant of the rigidity term support the stability 
of strings described by the effective action $S_{\rm eff.}$. 
While the requirement of positiveness of the string tension is 
obvious already for the very existence of strings, it is worth 
briefly discussing the requirement of the negativeness of the 
coupling constant of the rigidity term. A simple argument 
in favour of this observation can be obtained by considering the 
propagator corresponding to the string effective action~(\ref{streffact}) 
in the so-called conformal gauge for the induced metric, 
$\hat g^{ab}=\frac{\delta^{ab}}{\sqrt{\hat g}}$. 
For a certain Lorentz index $\lambda$ it reads 

$$
\left<x_\lambda(\xi)x_\lambda(0)\right>=
\int\frac{d^2p}{(2\pi)^2}\frac{{\rm e}^{ip^a\xi_a}}{\sigma p^2-
\frac{1}{\alpha_0}(p^2)^2}.$$
For negative $\alpha_0$, this integral is well defined and 
is equal to 

$$
\left<x_\lambda(\xi)x_\lambda(0)\right>=-\frac{1}{2\pi\sigma}
\left[\ln(\mu|\xi|)+K_0\left(\sqrt{\left|\alpha_0\right|\sigma}
|\xi|\right)\right],$$
where $\mu$ denotes the IR momentum cutoff, and $K_0$ is the modified 
Bessel function. Contrary to that, for positive $\alpha_0$ an 
unphysical pole 
in the propagator occurs, which confirms our statement.

Thus we conclude that the obtained string 
characteristica manifest confinement and provide us 
with the necessary condition for the stability of strings in the 
obtained effective Abelian-projected theory~(\ref{13}). 

\section{Summary and Discussions}

In the present paper, 
by considering a grand canonical ensemble of fluctuating 
monopole-like excitations emerging in the Abelian-projected 
$SU(2)$-gluodynamics, we have derived an effective disorder field 
theory describing this ensemble in the continuum limit. 
Contrary to the previous approaches, 
this has been done without an assumption on the formation and condensation of 
Cooper pairs of monopoles, {\it i.e.} without introducing the  
corresponding magnetic Higgs 
field. Instead of that, 
we have dealt directly with the 
dilute Coulomb gas of monopole loop currents 
(describing the creation and annihilation 
of the monopole-antimonopole pairs). 
The proposed approach provided 
us with a natural dynamical mechanism of generation of a mass of the 
dual gauge field, which is due to the Debye screening in the 
gas of monopole loop currents. 

Next, within the obtained theory 
we have investigated  the string representation 
of the Wilson loop, which describes an external particle electrically 
charged {\it w.r.t.} the maximal Abelian $U(1)$-subgroup of the 
original $SU(2)$-group. The essence 
of this representation is a certain mechanism realizing the 
independence of the Wilson loop of some surface, bounded by its 
contour. As it has been illustrated, this mechanism is based 
on the summation over 
branches of the multivalued effective potential of monopole 
loop currents. Finally, 
in the weak-field 
limit of the obtained effective Abelian-projected theory, 
we have derived the string tension of the Nambu-Goto term and the 
inverse coupling constant of the rigidity term, which in a 
manifest way express confinement in the sense of the Wilson's 
area law and signal the 
stability of strings.
In particular, the string tension turned out 
to be nonanalytic in the QCD coupling constant analogously to what 
happens in the original gluodynamics.  

In conclusion,
it is worth mentioning that the approach to the problem of 
confinement, investigated in the present 
work, essentially employed the 
disorder (stochastic) nature of fluctuations of monopole loop 
currents and related
Dirac strings. In particular, the resulting stochastic 
correlations of topological monopole-like excitations 
are an important dynamical 
ingredient for getting the Wilson's area 
law. Note that this dynamical mechanism confines (chromo)electric charges  
of both test quarks and gluons. In this sense, the above disorder 
(stochastic) approach differs from the standard one, 
where the area law follows from the nontrivial linking of some 
topological objects (like $Z_N$-vortices~\cite{eza}) 
with the contour of the Wilson loop. 
Notice that if one had fixed the gauge further, leaving, for example, 
the $Z_N$ center symmetry, the gluons would have been uncharged, and the 
physics of their confinement would be left obscure~\cite{kron}. 
Note also that, contrary to the most well known monopoles, 
Abelian-projected monopoles are described by 
stochastic loop 
currents. Owing to that, associated stochastically distributed 
Abrikosov-Nielsen-Olesen type 
strings neither do not need
to be a solution of the usual classical field equations of motion nor to be 
described by a vacuum expectation value of some field.

Clearly, it is now a challenge to apply the present approach 
to the more realistic case of $SU(3)$-gluodynamics. 
Work in this direction is now in progress~\cite{prep}.

\section*{Acknowledgments}

We are  indebted to  A. Di Giacomo, H.G. Dosch, 
J. Ellis, N.E. Mavromatos, H. Reinhardt, 
and Yu.A. Simonov for fruitful and 
illuminating discussions. 
D.E. expresses his thanks to H. Reinhardt 
for kind hospitality during his stay at the 
University of T\"ubingen and to the colleagues 
of the Theory Division at CERN for their hospitality
during another stay, where the paper reached its final form.
D.A. acknowledges 
the Quantum Field 
Theory Divisions of the Institute of Physics of the Humboldt 
University of Berlin and of the University of Pisa 
for cordial hospitality and the Graduate College 
{\it Elementarteilchenphysik} of the Humboldt University and INFN 
for financial support.


\newpage


\begin{thebibliography}{300}

\bibitem{th}
G. 't Hooft, Nucl. Phys. {\bf B 190} (1981) 455.

\bibitem{2}
S. Mandelstam, Phys. Lett. {\bf B 53} (1975) 476; 
Phys. Rep. {\bf C 23} (1976) 245; G. 't Hooft, in: 
{\it High Energy Physics}, Ed. A. Zichichi (Editrice Compositori, 
Bologna, 1976).

\bibitem{3}
A. Di Giacomo, preprint {\tt hep-lat/9907010} (1999); 
preprint {\tt hep-lat/9907029} (1999).

\bibitem{rein}
M. Quandt and H. Reinhardt, Int. J. Mod. Phys. {\bf A 13} (1998) 4049; 
Phys. Lett. {\bf B 424} (1998) 115.


\bibitem{kond}
K.-I. Kondo, Phys. Rev. {\bf D 57} (1998) 7467.

\bibitem{suz}
T. Suzuki, Prog. Theor. Phys. {\bf 80} (1988) 929.


\bibitem{ano}
A.A. Abrikosov, Sov. Phys.- JETP {\bf 5} (1957) 1174; 
H.B. Nielsen and P. Olesen, Nucl. Phys. {\bf B 61} (1973) 45;
for a review see {\it e.g.} E.M. Lifshitz and L.P. Pitaevski, 
{\it Statistical Physics, Vol. 2} (Pergamon, New York, 1987).


\bibitem{lee}
K. Lee, Phys. Rev. {\bf D 48} (1993) 2493; P. Orland, Nucl. Phys. 
{\bf B 428} (1994) 221; M. Sato and S. Yahikozawa, Nucl. Phys. 
{\bf B 436} (1995) 100; 
M. Kiometzis {\it et al.}, Fortschr. Phys. 
{\bf 43} (1995) 697.


\bibitem{polikarp}
E.T. Akhmedov {\it et al.}, Phys. Rev. {\bf D 53} (1996) 2087.

\bibitem{eur}
D. Antonov and D. Ebert, Eur. Phys. J. {\bf C 8} (1999) 343; 
in: {\it Problems of Quantum Field Theory}, Eds. B.M. Barbashov 
{\it et al.} (JINR, Dubna, 1999), pp. 285-290.

\bibitem{more}
D. Antonov and D. Ebert,  Phys. Lett. {\bf B 444} (1998) 208; 
D.A. Komarov and M.N. Chernodub, JETP Lett. {\bf 68} (1998) 117; 
D. Antonov and D. Ebert, in: {\it Path Integrals from peV 
to TeV: 50 years after Feynman's paper}, 
Eds. R. Casalbuoni {\it et al.}
(World Scientific, Singapore, 1999), 
pp. 267-271; preprint {\tt CERN-TH/99-294}, {\tt hep-th/9909156} (1999) 
(Nucl. Phys. {\bf B} (Proc. Suppl.), in press).  

\bibitem{bramb}
M. Baker {\it et al.}, Phys. Rev. {\bf D 58} (1998) 034010; 
U. Ellwanger, Eur. Phys. J. {\bf C 7} (1999) 673. 
 

\bibitem{mpla}
D.V. Antonov, D. Ebert, and Yu.A. Simonov, Mod. Phys. Lett. {\bf A 11} 
(1996) 1905; D.V. Antonov and D. Ebert, Mod. Phys. Lett. {\bf A 12} 
(1997) 2047; Phys. Rev. {\bf D 58} (1998) 067901.

\bibitem{rid}
A.M. Polyakov, Nucl. Phys. {\bf B 268} (1986) 406; 
H. Kleinert, Phys. Lett. {\bf B 174} (1986) 335. 


\bibitem{wil}
K.G. Wilson, Phys. Rev. {\bf D 10} (1974) 2445.

\bibitem{dec}
D. Antonov and D. Ebert, preprint {\tt 
HUB-EP-98/73, hep-th/9812112} (1998) (Eur. Phys. J. {\bf C}, in press).


\bibitem{confstr}
A.M. Polyakov, Nucl. Phys. {\bf B 486} (1997) 23.



\bibitem{dia}
M.C. Diamantini, F. Quevedo, and C.A. Trugenberger, Phys. Lett. 
{\bf B 396} (1997) 115.



\bibitem{fuc}
F. Fucito, M. Martellini, and M. Zeni, Nucl. Phys. {\bf B 496} (1997) 259.


\bibitem{bilocal}
H.G. Dosch, Phys. Lett. {\bf B 190} (1987) 177; 
Yu.A. Simonov, Nucl. Phys. {\bf B 307} (1988) 512;
H.G. Dosch, Prog. Part. Nucl. Phys. {\bf 33} (1994) 121;
Yu.A. Simonov, Phys. Usp. {\bf 39} (1996) 313.



\bibitem{backgr}
B.S. de Witt, Phys. Rev. {\bf 162} (1967) 1195, 1239; 
J. Honerkamp, Nucl. Phys. {\bf B 48} (1972) 269; 
G. 't Hooft, Nucl. Phys. {\bf B 62} (1973) 444; 
L.F. Abbot, Nucl. Phys. {\bf B 185} (1981) 189. 

\bibitem{ellw}
U. Ellwanger, preprint {\tt hep-th/9906061} (1999).


\bibitem{rein1}
H. Reinhardt, Nucl. Phys. {\bf B 503} (1997) 505.


\bibitem{abdom}
Z.F. Ezawa and A. Iwazaki, Phys. Rev. {\bf D 25} (1982) 2681; 
ibid. {\bf D 26} (1982) 631.

\bibitem{KR}
M. Kalb and P. Ramond, Phys. Rev. {\bf D 9} (1974) 2237.

\bibitem{eza}
Z.F. Ezawa and A. Iwazaki, Phys. Rev. {\bf D 23} (1981) 3036; 
ibid. {\bf D 24} (1981) 2264.


\bibitem{kron}
A.S. Kronfeld, G. Schierholz, and U.-J. Wiese, Nucl. Phys. {\bf B 293} 
(1987) 461.

\bibitem{prep}
D. Antonov and D. Ebert, in preparation. 

\end{thebibliography}
\end{document}